\documentclass[preprint,onecolumn,nofootinbib]{revtex4}
\usepackage[colorlinks=true,linkcolor=blue,urlcolor=blue,filecolor=black,citecolor=red,pdfstartview=FitV,pdftitle={},pdfsubject={},pdfkeywords={},pdfpagemode=None,bookmarksopen=true]{hyperref}
\usepackage{autobreak}
\usepackage{graphicx}
\usepackage{amsmath}
\usepackage{amsfonts}
\usepackage{amssymb,ulem}
\usepackage{color}%
\usepackage{tikz}
\usepackage{dcolumn}
\usepackage{enumerate}
\usepackage{xcolor}
\usepackage{tabularray}

\definecolor{lightblue}{rgb}{0.8, 0.9, 1}
\definecolor{lightred}{rgb}{1, 0.9, 0.8}
\definecolor{lightgreen}{rgb}{0.8, 1, 0.8}

\allowdisplaybreaks

\begin{document}
\title{
    Diagnosing Emergent Isotropy in Anisotropic Holographic Systems using Quantum Information Measures
}

\author{Chong-Ye Chen $^{1}$}
\email{cychen@stu2022.jnu.edu.cn}
\author{Mu-Jing Li $^{1}$}
\email{mujing.phy@gmail.com}
\author{Zhe Yang $^{1}$}
\email{yzar55@stu2021.jnu.edu.cn}
\author{Da-Ming Jin $^{1}$}
\email{Drop39@outlook.com}
\author{Peng Liu $^{1}$}
\email{phylp@email.jnu.edu.cn}
\thanks{corresponding author}
\affiliation{
    $^1$ Department of Physics and Siyuan Laboratory, Jinan University, Guangzhou 510632, China
}

\begin{abstract}

    This study presents a comprehensive investigation of anisotropy in a holographic p-wave superconductor model, revealing novel insights into the behavior of quantum information measures in strongly coupled systems. Through rigorous semi-analytical methods, we uncover the existence of an isotropic point emerging at a critical temperature $T_{II}$, marking a significant transition in the system's anisotropic properties. We offer a systematic analysis of the mechanisms driving anisotropy and isotropy transitions, finding that this phenomenon is unique to the p-wave model and absent in other anisotropic systems like anisotropic axion models with metal-insulator transitions. We propose that the explicit component dependence of the vector field manifesting anisotropy is the key driver of the emergent isotropy. Our analysis of holographic entanglement entropy (HEE), entanglement wedge cross-section (EWCS), and butterfly velocity demonstrates their distinct sensitivities to bulk anisotropy, with EWCS and butterfly velocity emerging as superior probes for detecting the isotropic point.
    
\end{abstract}
\maketitle
\tableofcontents

\section{INTRODUCTION}

Natural systems exhibit anisotropy across a broad spectrum, from Weyl semimetals to magnetic structures \cite{Ma:2021,Li:2021,Rohrer:2011,Berkowitz:1999,Kok:2018,Song:1997,Quirk:2023,Glotzer:2007}. Many phase transitions are notably marked by the transition from isotropic to anisotropic states, a key phenomenon in the study of quantum materials. This development of anisotropy is pivotal, as it indicates the loss of global symmetry and confers direction-dependent properties in the realms of electronics, magnetism, and optics. The implications are significant, affecting electronic transport, the flow of information, and altering topological characteristics. Moreover, anisotropy bears a critical relationship with entanglement measures in certain strongly correlated systems and serves as an innovative asset in precision measurement techniques. The investigation into anisotropy's effect on entanglement characteristics remains an important but formidable field of study, primarily due to the dual complexities associated with strongly correlated systems and the intricate nature of calculating entanglement measures, both of which are notoriously challenging to compute \cite{Liu:2019npm,Amico:2007ag, Laflorencie:2016,Horodecki:2009zz}.

In the context of the AdS/CFT (Anti-de Sitter/Conformal Field Theory) correspondence—a crucial tool in theoretical physics that provides a duality between a gravity theory in a higher-dimensional space (AdS) and a strongly correlated field theory on its boundary—the anisotropy of the boundary theory can be encoded in the geometry of the bulk gravitational theory \cite{Maldacena:1997re, Gubser:1998bc, Hubeny:2007xt, Lewkowycz:2013nqa}. This deep connection allows for the study of quantum information in strongly correlated systems through the lens of a gravitational theory. In the holographic duality, researchers have made progress in understanding quantum information, that can be broadly categorized into two categories: properties encoded in a slice of space that does not change over time (static) and properties encoded in dynamic processes (dynamic). 

Static quantum information includes prominent measures such as the holographic entanglement entropy (HEE) and the entanglement wedge cross-section (EWCS). The HEE, governed by the Ryu-Takayanagi formula, links the entanglement in a boundary CFT to the geometry of a gravitational spacetime via minimum surfaces \cite{Ryu:2006bv, Ryu:2006ef, Nishioka:2006gr, Klebanov:2007ws, Pakman:2008ui}. The EWCS extends this by offering a holographic dual of mixed-state correlations that have been associated with the entanglement of purification, logarithmic negativity and reflected entropy \cite{Terhal:2002zz, Vidal:2002zz, Plenio:2005cwa, Caputa:2018xuf, Liu:2019qje, Takayanagi:2017knl, Nguyen:2017yqw, Kudler-Flam:2018qjo, Liu:2020blk, Chen:2021bjt, Cheng:2021hbw}. 

The dynamic one, revolves around inherently time-dependent properties, focusing on how quantum information evolves or behaves through dynamic processes. It targets understanding how quantum information transitions or flows through various states, particularly focusing on the mechanisms behind the spreading and evolution of entanglement. An essential feature under this dynamic category is the concept of the butterfly velocity. This velocity is a critical measure for gauging the rate at which quantum correlations propagate across a strongly entangled system, which is intricately tied to the concept of quantum scrambling and thermalization detected using out-of-time-order correlators (OTOCs) \cite{Blake:2016wvh,Ling:2017jik,Baggioli:2018afg,Roberts:2014isa,Maldacena:2015waa,Polchinski:2015cea,Roberts:2016wdl}. The interplay between anisotropy and quantum information is a significant theme that permeates both the static and dynamic aspects of these holographic measures.

Anisotropy in strongly correlated systems influences quantum information properties, which are accessible through the AdS/CFT correspondence. This influence can be explored by examining static measures such as HEE and EWCS, as well as dynamic measures like butterfly velocity, in a holographic system with anisotropy. This raises the question: what types of systems can show anisotropy in a holographic framework?

Constructing a holographic system with anisotropy typically involves introducing mechanisms that break the isotropy. A direct way to do this is by adding matter fields with anisotropic stress-energy tensors to the gravitational theory which can result in anisotropic geometries. These matter fields could be scalar fields, vector fields, or higher-form fields with nontrivial profiles along specific spatial directions. For example, anisotropic axions or anisotropic lattices can be introduced to break the isotropy \cite{Ling:2015ghh,Fang:2014jka,Arefeva:2018hyo,Donos:2013gda,Ling:2014saa}. Another way to break the isotropy is to incorporate a field that breaks the isotropy more explicitly, such as vector fields along certain directions or anisotropic tensor condensations. One such example is the p-wave superconductor model \cite{Cai:2012nm,Cai:2013oma,Cai:2013aca,Cai:2014ija,Yang:2023wuw,Cai:2021obq,Li:2013rhw}
\footnote{The proof that p-wave model under the Einstein-SU(2) Yang-Mills theory \cite{Cai:2012nm,Cai:2013oma} is a special case under the Einstein-Maxwell theory with a complex vector field \cite{Cai:2013aca,Li:2013rhw,Cai:2014ija} can be found in \cite{Cai:2021obq}.}.
In this model, the vector field breaks the isotropy, and the resultant anisotropy can be enhanced by tuning the parameters of the model. Previous studies on anisotropic effects on entanglement-related quantities can be found in \cite{Ahn:2017kvc,Jahnke:2017iwi,Avila:2018sqf,Jokela:2019ebz,Jokela:2019tsb,Dudal:2018ztm,Dey:2014voa,Mishra:2016yor,Mishra:2018tzj,Gursoy:2018ydr,Roychowdhury:2015fxf,Mahapatra:2019uql,Narayan:2012ks,Narayan:2013qga,Mukherjee:2014gia,Narayan:2015lka,Giataganas:2012zy,Giataganas:2013lga,Vasli:2022kfu,Gursoy:2020kjd,Giataganas:2017koz,Chu:2019uoh,Baggioli:2020cld}. In this work, we investigate the anisotropy in a holographic p-wave model. As the temperature decreases below a critical point, an isotropic point emerges in the infrared (IR) region. We conduct a thorough analysis of the anisotropy and the emergence of isotropic points, and subsequently explore the anisotropy in quantum information measures, providing a comprehensive understanding of their behavior within this system.

This paper is organized into four parts: In Section \ref{sec:model}, we introduce the p-wave superconductor model and analyze its anisotropy. Section \ref{sec:QI} introduces three anisotropic information-related quantities. In Section \ref{sec:computation}, we study anisotropic HEE, EWCS, and the butterfly velocity. Finally, in Section \ref{sec:discuss}, we summarize our findings and conclusions.

\section{The Holographic p-wave Superconductor Model and Analysis on Anisotropy}\label{sec:model}

\subsection{The Holographic p-wave Superconductor Model}\label{sec:model2}

In holographic superconductor, the symmetry of the system is spontaneously broken as the temperature decreases below the critical point, leading to the emergence of the superconducting phase \cite{Hartnoll:2008kx, Horowitz:2010gk, Hartnoll:2008vx}. Recent studies have delved into various types of holographic superconductor theories, with one particularly notable example being the holographic p-wave superconductor \cite{Cai:2010cv, Gangopadhyay:2012gx,Cai:2012nm,Cai:2013oma,Cai:2013aca,Cai:2014ija,Yang:2023wuw,Cai:2021obq,Li:2013rhw}. The p-wave superconductor possesses anisotropy, and in the framework of holographic duality, a holographic p-wave superconductor must also manifest this anisotropy. In this paper, we study a holographic p-wave superconductor theory incorporating a complex vector field into the Einstein-Maxwell theory with a negative cosmological constant \cite{Cai:2013aca,Li:2013rhw, Cai:2014ija}. The Lagrangian density for this theory is given by:
\begin{equation} \label{Ld}
    \mathcal{L}= \sqrt{-g}\left(\mathcal{R}+\frac{6}{L^{2}}-\frac{1}{4} F_{\mu \nu} F^{\mu \nu}-\frac{1}{2} \rho_{\mu \nu}^{\dagger} \rho^{\mu \nu}-m^{2} \rho_{\mu}^{\dagger} \rho^{\mu}+i q \gamma \rho_{\mu} \rho_{\nu}^{\dagger} F^{\mu \nu}\right),
\end{equation}
where $L$ represents the AdS radius and is set to 1 for concreteness. The field strength tensor is given by $F_{\mu\nu} = \nabla_\mu A_\nu - \nabla_\nu A_\mu$, where $A_\mu$ represents the Maxwell field. $\rho_\mu$ is a complex vector field, which possesses mass $m$ and charge $q$. The field strength tensor for $\rho_\mu$ is defined as $\rho_{\mu\nu} = D_\mu \rho_\nu - D_\nu \rho_\mu$, where the covariant derivative $D_\mu$ is defined as $D_\mu = \nabla_\mu - i q A_\mu$. Note that the non-minimal coupling between the complex vector field and the Maxwell field can be turned off by setting $\gamma$ to zero, which is appropriate in the absence of the magnetic field in our current context.

The equations of motion from \eqref{Ld} are,
\begin{equation} \label{eom1}
    \begin{aligned}
        \nabla^{\nu} F_{\nu \mu} & = i q\left(\rho^{\nu} \rho_{\nu \mu}^{\dagger}-\rho^{\nu \dagger} \rho_{\nu \mu}\right) + i q \gamma \nabla^{\nu}\left(\rho_{\nu} \rho_{\mu}^{\dagger}-\rho_{\nu}^{\dagger} \rho_{\mu}\right),                                                                                              \\
        D^{\nu} \rho_{\nu \mu}   & = m^{2} \rho_{\mu} - i q \gamma \rho^{\nu} F_{\nu \mu},                                                                                                                                                                                                                                     \\
        \mathcal{R}_{\mu \nu}    & = \frac{1}{2} \mathcal{R} g_{\mu \nu} + \frac{3}{L^{2}} g_{\mu \nu} + \frac{1}{2} F_{\mu \lambda} F_{\nu}{}^{\lambda}                                                                                                                                                                       \\
                                 & \quad + \frac{1}{2}\left(-\frac{1}{4} F_{\mu \nu} F^{\mu \nu} - \frac{1}{2} \rho_{\mu \nu}^{\dagger} \rho^{\mu \nu} - m^{2} \rho_{\mu}^{\dagger} \rho^{\mu} + i q \gamma \rho_{\mu} \rho_{\nu}^{\dagger} F^{\mu \nu}\right) g_{\mu \nu}                                                     \\
                                 & \quad + \frac{1}{2}\left\{\left[\rho_{\mu \lambda}^{\dagger} \rho_{\nu}^{\lambda} + m^{2} \rho_{\mu}^{\dagger} \rho_{\nu} - i q \gamma\left(\rho_{\mu} \rho_{\lambda}^{\dagger} - \rho_{\mu}^{\dagger} \rho_{\lambda}\right) F_{\nu}{}^{\lambda}\right] + \mu \leftrightarrow \nu\right\}.
    \end{aligned}
\end{equation}
We consider the black brane solution with asymptotic AdS spacetime, which takes the following ansatz,
\begin{equation}\label{metric}
    \begin{aligned}
        d s^{2}           & =\frac{1}{z^{2}}\left(-p(z)(1-z) U(z) d t^{2}+\frac{1}{p(z)(1-z) U(z)} d z^{2}+V_{1}(z) d x^{2}+V_{2}(z) d y^{2}\right) \\
        A_{\nu} d x^{\nu} & =\mu(1-z) a(z) d t, \quad \rho_{\nu} d x^{\nu}=\rho_{x}(z) d x
    \end{aligned}
\end{equation}
where $p(z) \equiv 1+z+z^{2}-\frac{\mu^{2} z^{3}}{4}$, and $\mu$ is the chemical potential of the dual field theory. This compact coordinate system designates the black brane horizon at $z=1$, while the asymptotic boundary of the spacetime is situated at $z=0$. We set boundary conditions $U(0)=V_1(0)=V_2(0)=1$ to assure the asymptotic AdS. The ansatz \eqref{metric} reduces to the AdS-RN black brane solution when the vector hair $\rho_{\mu}$ vanishes, and $U = V_1 = V_2 = a = 1$. Furthermore, by setting $U(1)=1$, the Hawking temperature of this black brane is given by $\tilde{T}=\frac{12-\mu^2}{16 \pi}$. We are interested in scaling-invariant physical quantities, therefore, we adopt the chemical potential $\mu$ as the scaling unit. As a result, the dimensionless Hawking temperature is $T=\tilde{T} / \mu$.

This holographic p-wave superconductor model exhibits a variety of phase transitions, including zeroth-order, first-order, and second-order transitions \cite{Cai:2013aca}. Our focus in this article is on a specific set of superconducting phases below the critical temperature $T_c$. A chosen action parameters, $q=1.5$ and $m^2=3/4$, are representative, since other choices will show similar anisotropic phenomena, as discussed in section \ref{sec:model4}. In this case, these black branes experience a second-order phase transition at $T_c = 0.0126$.

\subsection{The Analysis on Anisotropy in Holographic P-wave Superconductor Model}\label{sec:model3}

Tuning down the temperature below the critical point of phase transition $T_c$, we get a black brane solution with a non-vanishing vector field $\rho_{\mu}$. Recently, the behavior of the condensate $\langle J_x \rangle$ in the vicinity of this critical point has been examined \cite{Yang:2023wuw}. The presence of a non-trivial $x$ component $\rho_x$ in the vector field leads to an anisotropy in the $x$-$y$ plane. This anisotropy is represented explicitly by the inequality $V_1 \neq V_2$, as illustrated in Fig. \ref{fig:V1V2q15}\footnote{Anisotropy is a term with broad implications, indicating variations in properties or behaviors depending on direction. When it comes to gravitational systems, this concept can be seen in the geometry of space or in the characteristics of matter. However, our primary focus is on background anisotropy—those directional variations that are built into the very framework of a system. We concentrate on the anisotropy of the background because they fundamentally guide how every other aspect of the system operates.}. Although the background solution of the gravity system exhibits anisotropy in the $x$-$y$ plane, it is important to note that the background solutions remain homogeneous. This means that all functions, including $ U, V_1, V_2, a, \rho_x $, are only dependent on the $ z $-coordinate.

To clarify the origin of the anisotropy in this system, it is essential to highlight the distinctive role of the vector field components. In this context, the anisotropy is pronounced by the fact that $\rho_x$ is non-zero, while $\rho_y$ is precisely zero. This discrepancy is not only reflected in the dynamics of the system but also in the very structure of the potential term in the action, which includes $\rho_x$ exclusively,
\begin{equation} 
    -m^2 \rho^\dagger_\mu \rho^\mu = - \frac{m^2 z^2 \rho_x(z)^2}{V_1(z)}.
\end{equation}
This potential term indicates anisotropy, in which $\rho_x$ is coupled only to $V_1$, the metric component in the $x$-direction—a stark contrast to the isotropic behavior observed in $s$-wave scenarios. Moreover, the field equations provide a compelling argument for this anisotropy. The vector field strength $\rho_{\mu\nu}$ plays a significant role in the $T_{xx}$ component of the energy-momentum tensor, contributing to the momentum flux along the $x$-axis.
\begin{equation}
    \begin{aligned}
        (\rho^\dagger_{x\lambda}\rho^\dagger_{x}{}^\lambda+m^2 \rho^\dagger_x\rho^\dagger_x)= & \left( m^2-\frac{4q^2(-1+z)z^2\mu^2 a(z)^2}{(-4(1+z+z^2)+z^3\mu^2)U(z)} \right)\rho_x(z)^2 \\
                                                                                              & +\frac{1}{4}z^2(4+z^3(-4+(-1+z)\mu^2))U(z)\rho_x'(z)^2
    \end{aligned}
\end{equation}
However, it is entirely absent in the case of the $T_{yy}$ component, indicating no such flux in the perpendicular $y$-axis direction. The anisotropy is thus rooted in the unique involvement of $\rho_x$ in action and equations of motion (EoMs).

Next, in order to acquire an in-depth understanding of anisotropy, we employ a semi-analytical approach for its examination.

\subsection{The Semi-Analytical Analysis on the Existence of Isotropic Point in Holographic P-wave Superconductor Model}\label{sec:model4}

In this section, we delve into a semi-analytical analysis of the holographic p-wave superconductor model, focusing on the existence of an isotropic point in the superconducting phase. We begin by showcasing numerical simulations that illustrate the change of anisotropy and the emergence of an isotropic point. Subsequently, we seek to understand these numerical results through a semi-analytical approach.

\begin{figure}
    \centering
    \includegraphics[height=0.22\textwidth]{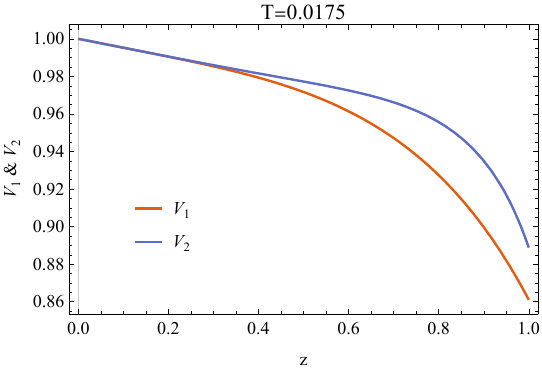}
    \includegraphics[height=0.22\textwidth]{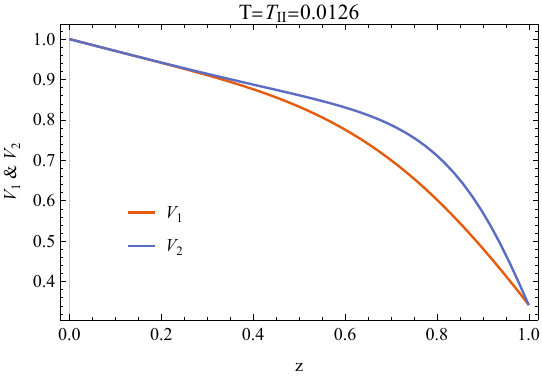}
    \includegraphics[height=0.22\textwidth]{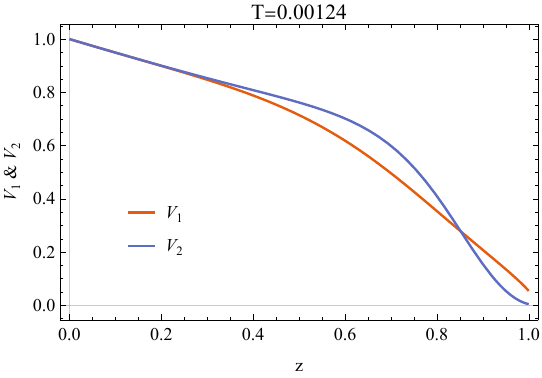}
    \caption{$V_{1}(z)$ and $V_{2}(z)$ profiles for $T>T_{II}$, $T_{II}$ and $T<T_{II}$ background solutions.}
    \label{fig:V1V2q15}
\end{figure}

\begin{figure}
    \centering
    \includegraphics[height=0.4\textwidth]{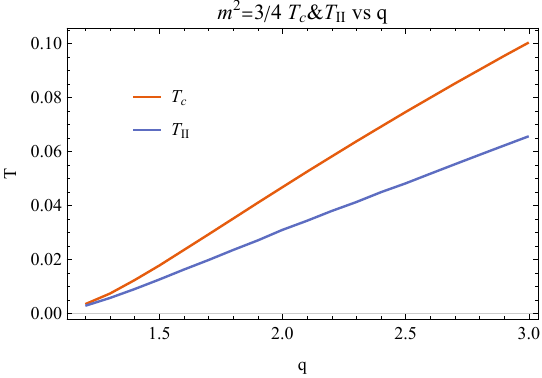}
    \caption{$T_c$ and $T_{II}$ vs $q$ with $m^2 = 3/4$.}
    \label{fig:TcTII}
\end{figure}

Numerical simulations allow us to study how anisotropy evolves as a function of the radial coordinate $ z $, which corresponds to the energy scale in the dual field theory. The change of anisotropy with the decreasing temperature is shown from left to right in FIG. \ref{fig:V1V2q15}. In this process, the decreasing temperature further enhances the emergent anisotropy of the system. In addition to this anisotropy, the background geometry reveals a significant phenomenon at a critical temperature where $V_1(1)=V_2(1)$.
This marks a transition point or a special scale where isotropy emerges, potentially signifying symmetry restoration or interesting dynamics in the field theory at that scale. We term this temperature as the isotropic temperature of IR, denoted as $T_{II}$. Furthermore, we numerically compute $T_c$ and $T_{II}$ for a range of $q$, in which $T_{II}$ is consistently smaller than $T_{c}$ (as shown in FIG. \ref{fig:TcTII}). 
As the temperature decreases further, the isotropic point shifts from the scale $z=1$ (the horizon) into the scale $0<z<1$ (the bulk), as indicated by the process from $V_1(1) < V_2(1)$ to $V_1(1) > V_2(1)$ (as shown in Fig. \ref{fig:V1V2q15}). This shift signifies a substantial alteration in the system's dynamics associated with anisotropic properties near the spacetime horizon. In addition to this emerging isotropic point in geometry, a related effect is observed in the energy-momentum tensor, where $T_{xx}=T_{yy}$ at a certain scale $z_m$.\footnote{Due to the nonlinear coupling of $V_1$ and $V_2$ in the Einstein equations, the isotropic point does not necessarily coincide with $z_m$.} These behaviors represent a novel and intriguing observation different from those observed in many other anisotropic models. The technical details for the numerical solution can be found in \ref{app:technical}. Now, we proceed with an analysis of the existence of the isotropic point and the transition of anisotropy near the horizon.

We study the analytical understanding of anisotropy from two parameters, encompassing energy scales at the ultraviolet (UV) and infrared (IR) boundaries, alongside a temperature axis that spans from the critical temperature $ T_c $ to absolute zero, in order to decipher the underlying physics of a p-wave superconductor's behavior. At the UV and IR limits, the boundary conditions and asymptotic limits allow for insightful analytical signals about the anisotropic nature of spacetime itself. Near $ T_c $, the p-wave phase transition provides fertile ground for perturbative methods, granting access to insights into how anisotropy emerges and evolves near this critical point. As we approach the zero-temperature limit, a similar analytical approach becomes available, unmasking the properties of the superconducting state in the absence of thermal fluctuations, where quantum effects preside.

In the following semi-analytical analysis, we focus on the anisotropy of four regions, that is, the UV and IR regions below but close to the critical point $T \lesssim T_c$, and the UV and IR regions in the limit of zero temperature $T \to 0$. By studying the anisotropy of these four regions, we can confirm the existence of the isotropic point within the bulk of the spacetime.
First (for UV region at $T \lesssim T_c$), to investigate the influence of the vector field $\rho_\mu$ on the spacetime dynamics near the critical temperature $T \lesssim T_c$, we consider the following perturbation \cite{Pan:2012jf},
\begin{equation} \label{perturbation}
    \begin{aligned}
        U(z)      & = 1+\epsilon^2 U^{(2)}(z) + \epsilon^4 U^{(4)}(z) \cdots       \\
        V_1(z)    & = 1+\epsilon^2 V_1^{(2)}(z)+ \epsilon^4 V_1^{(4)}(z) \cdots    \\
        V_2(z)    & = 1+\epsilon^2 V_2^{(2)}(z)+ \epsilon^4 V_2^{(4)}(z) \cdots    \\
        a(z)      & = 1+ \epsilon^2 a^{(2)}(z) +\epsilon^4 a^{(4)}(z) \cdots       \\
        \rho_x(z) & = \epsilon \rho_x^{(1)}(z)+ \epsilon^3 \rho_x^{(3)}(z) \cdots
    \end{aligned}
\end{equation}
where we are especially interested in examining the impact of the presence of a vector field on anisotropy at the leading order. Plugging \eqref{perturbation} into the equations of motion (EoMs) and taking $z\to 0$ limit, we obtain
\begin{equation} \label{perturbres1}
    \begin{aligned}
         & V_1^{(2)}( z)-V_2^{(2)}(z) \approx - 2 q^2 \rho^{(1)}_x (z)^2 <0.
    \end{aligned}
\end{equation}
The obtained results indicate that near the UV region, $V_1(z)$ is qualitatively smaller than $ V_2(z) $. This qualitative alignment with our numerical solutions further supports the validity and consistency of our results.

Similarly (for IR region at $T \lesssim T_c$), taking $z\to 1$ limit in the leading order of the perturbation equations, we can obtain the following results,
\begin{equation} \label{perturbres2}
    \begin{aligned}
         & V_1^{(2)}{}'(z) - V_2^{(2)}{}'(z)  \approx \frac{8m^2\rho_x^{(1)}(z)^2}{12-\mu^2} > 0,
    \end{aligned}
\end{equation}
where the chemical potential $\mu^2<12$ at the finite temperature. This indicates that $V_1'(z)$ is qualitatively larger than $V_2'(z)$, which is consistent with our numerical results. The qualitative behavior of $V_1(z)-V_2(z)$ near the IR region can be determined by solving the sub-leading order perturbation equations. These equations involve additional sub-leading perturbative terms, which require more boundary conditions to determine the behavior. However, the boundary conditions at the horizon are insufficient to achieve this. The qualitative behavior of $V_1(z)-V_2(z)$ near the IR region requires a smooth interplay between the UV region (with boundary conditions for the AdS) and the IR region (with boundary conditions for the horizon). Therefore, a numerical method is needed to achieve this goal. According to the numerical solutions, $V_1(z)$ is smaller than $V_2(z)$ near the IR region at $T \lesssim T_c$. 

In the second step (for UV region at $T \to 0 $), we perform an asymptotic expansion near the AdS boundary in zero temperature limit. Throughout this analysis, we obtain the behavior of $V_1$ and $V_2$,
\begin{equation} \label{adsbdyres1}
    \begin{aligned}
         & V_1(0)=V_2(0)=1,\quad V_1'(0)=V_2'(0),\quad V_1''(0)=V_2''(0), \\
         & V_1'''(0)=-V_2'''(0), \quad V_1^{(4)}(0)=-V_2^{(4)}(0).
    \end{aligned}
\end{equation}
This suggests that $V_1(z) - V_2(z) \sim z^3$ in the UV region. Further information of the anisotropic behavior requires more boundary conditions. On the other hand, the numerical result reveals $V_1(z) - V_2(z) \sim z^5$ in the UV region. This require $V_1'''(0)=-V_2'''(0)=0,  V_1^{(4)}(0)=-V_2^{(4)}(0)=0$. After substituting these conditions into the expansion, we can obtain the following anisotropic behavior,
\begin{equation} \label{adsbdy2}
    \begin{aligned}
        V_1(z) - V_2(z) \approx -\frac{3}{20} \rho_x''(z)^2 z^5 < 0.
    \end{aligned}
\end{equation}
This indicates that $V_1(z)$ is qualitatively smaller than $V_2(z)$ in zero temperature limit. Additionally, this result is quantitatively verified by our numerical solutions.

\begin{figure}
    \centering
    \includegraphics[height=0.4\textwidth]{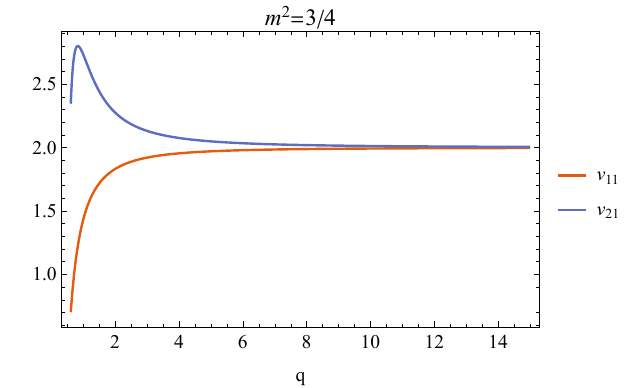}
    \caption{$v_{11}$ and $v_{21}$ as functions of $q$. This system always satisfies $v_{21}>v_{11}$ for any finite $q$.} 
    \label{fig:v11v21}
\end{figure}

In the last step (for IR region at $T \to 0$), we study the anisotropic behavior at the horizon in zero temperature limit, i.e. the anisotropy of the IR fixed points. We adopt the following setup,
\begin{equation} \label{rsetup}
    \begin{aligned}
         & ds^2=\left(-U(r)dt^2+\frac{1}{U(r)}dr^2+V_1(r)dx^2+V_2(r)dy^2\right), \\
         & A_\nu dx^\nu = a(r) dt,\quad \rho_\nu d x^\nu=\rho_x(r)dx.
    \end{aligned}
\end{equation}
To obtain the zero-temperature IR geometry, we propose such a power law solution for $r\to 0$,
\begin{equation} \label{rpower}
    \begin{aligned}
         & U(r) = u_{0} r^{u_{1}},\quad V_1(r) = v_{10} r^{v_{11}},\quad & V_2(r) = v_{20} r^{v_{21}} \\
         & a(r) = v_{20} r^{v_{21}},\quad \rho_x(r) = p_{0} r^{p_{1}}.
    \end{aligned}
\end{equation}
By substituting \eqref{rpower} into the EoMs, we can obtain the following algebraic equations,
\begin{equation} \label{IRres}
    \begin{aligned}
        v_{11} & =-\frac{\left(m^2-4q^2\right)v_{21}}{m^2+2q^2v_{21}} 
        \\
        0      & =-3\left(m^2-4q^2\right)v_{21}\left(2+v_{21}\right)\left(m^2+2q^2v_{21}\right)-\left(m^2+q^2(-2+v_{21})\right)\times \\ &\quad\quad \left(m^4\left(2+v_{21}+v_{21}{}^2\right)+2m^2q^2 v_{21}\left(6+v_{21}+v_{21}{}^2\right)+2q^4v_{21}{}^2\left(12+v_{21}\left(4+v_{21}\right)\right)\right),
    \end{aligned}
\end{equation}
where exponents $v_{11}$ and $v_{21}$ can be determined by fixing parameter $m$ and $q$. For the case discussed in this paper, the exponents of the IR geometry read,
\begin{equation} \label{IRres2}
    \begin{aligned}
         & v_{11}\approx 1.71634 ,\quad v_{21} \approx 2.44505,\quad ( m^2=3/4,q=3/2 )
    \end{aligned}
\end{equation}
Based on the above result, it is evident that as $r \to 0$, $V_2\sim r^{2.44505}$ vanishes more rapidly than $V_1\sim r^{1.71634}$. As a result, in the zero temperature limit, the anisotropic behavior $V_2 < V_1$ is observed in the IR region. Additionally, both $v_{11}$ and $v_{21}$ as functions of $q$ plotted in FIG. \ref{fig:v11v21} reveals a generality of this anisotropic behavior. 

Based on the above arguments, it can be inferred that below and close to the critical temperature $T\lesssim T_c$, both near the UV (AdS boundary) and the IR (event horizon), $V_1 $ is smaller than $ V_2$. However, as the temperature approaches zero $V_1 $ becomes greater than $ V_2$ in the IR region. Consequently, at a certain point $T_{II}<T_c$, an isotropic point emerges at the IR. Furthermore, according to Eq. \eqref{adsbdy2}, the anisotropy in the UV region $V_1 < V_2$ still holds even in the zero temperature limit. The above results qualitatively are independent of the value of $q$. We summarize the above results in Table. \ref{tab:combined}. According to this analysis, at least one isotropic point emerges as the temperature tends to zero. Furthermore, based on existing numerical results, there is only one isotropic point when $T<T_{II}$. In summary, the system undergoes a single transition from anisotropy to a state with one isotropic point as the temperature tends to zero. 

\begin{table}
    \centering
    \begin{tblr}{
        colspec      = {|>{\centering\arraybackslash}m{6cm}|*{6}{>{\centering\arraybackslash}m{3.2cm}|}},
        column {2} = {lightblue},
        column {3} = {lightred},
        }
        \hline
        Temperature/Region                     & UV $( z \to 0 )$                & IR $( z \to 1)$            \\ \hline\hline
        near critical point $(T \lesssim T_c)$ & $ V_1<V_2 $ (analytically)      & $V_1 < V_2$ (numerically)  \\ \hline
        zero-temperature $(T \to 0)$           & $V_1 < V_2$ (semi-analytically) & $V_1 > V_2$ (analytically) \\ \hline
        
    \end{tblr}
    \caption{
        Anisotropic behavior of the system in the UV (z $\to$ 0) and IR (z $\to$ 1) limits for near-critical (T $\lesssim$ T$_c$) and zero-temperature (T $\to$ 0) regions.
    }
    \label{tab:combined}
\end{table}

In the context of anisotropic models, the emergence of an isotropic point within this particular model is a distinctive and noteworthy characteristic. This feature distinguishes it from other anisotropic models that lack such an isotropic point and implies the presence of additional underlying physical mechanisms that govern the behavior of the system. We have also examined the anisotropic lattice model and anisotropic axion model \cite{Ling:2016wyr, Fu:2022qtz,Donos:2013eha,Donos:2014uba}, where isotropy is broken due to the lattice and axion deformations respectively. In these models, the explicit $x$ dependence of these deformations contributes to a momentum flux in the $x$ direction. Unlike the p-wave model, these models do not exhibit an emerging isotropic point into the bulk or IR region.
\footnote{We examine the boundary behavior of anisotropy in anisotropic axion model. Based on the IR geometry described in \cite{Donos:2014uba} and our numerical solutions for three representative cases \cite{Fu:2022qtz}, we find no evidence of an emerging isotropic point in this model. The technical details of our numerical analysis are provided in Appendix \ref{app:technical}.}

Therefore, the presence of an isotropic point in p-wave model is a distinct and unique feature. This observation raises our interest in understanding the impact of these isotropic points on the anisotropic behavior of holographic quantum information. Consequently, in the following sections, we will delve into the study of the anisotropic behavior of holographic quantum information within this model.

\section{The Staitc and Dynamical Holographic Information-related Quantities on anisotropic background}\label{sec:QI}

Having analyzed the anisotropy in the holographic p-wave superconductor model, we now introduce key information-related quantities that can characterize this anisotropic behavior. Specifically, we discuss computational methodologies to evaluate three crucial probes, encompassing static quantities such as HEE and EWCS, along with a dynamical quantity, the butterfly velocity.

\subsection{The Static Holographic Information-related Quantities}

Holographic entanglement entropy (HEE) is a concept emerging from the intersection of quantum gravity and quantum information theory, where the intricate connections between subregions of a quantum system are understood through the lens of holography. It suggests that the measure of entanglement between parts of a boundary theory can be represented geometrically by the surface area of a minimum surface within a higher-dimensional gravitational space \cite{Ryu:2006bv}.

\begin{figure}
    \includegraphics[height=0.4\textwidth]{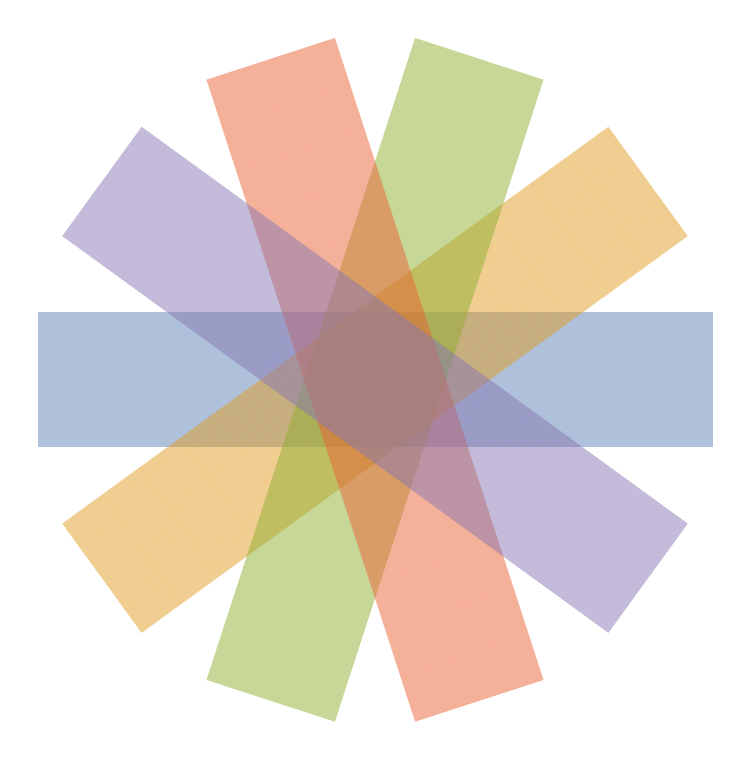}
    \caption{Visualization of strips oriented in five directions within the $x$-$y$ plane \cite{Liu:2019npm}.}
    \label{fig:plotshowrotate}
\end{figure}

Previous studies have predominantly investigated the behavior of HEE along a single direction, for instance, examining the HEE in the $x$ or $y$ direction within the context of p-wave models (see, e.g., \cite{Cai:2012nm,Cai:2013oma,Li:2013rhw,Yang:2023wuw}). Consequently, it becomes imperative to delve into the anisotropy on HEE in the anisotropic models. Employing computational techniques in \cite{Liu:2019qje,Liu:2019npm}, we investigate HEE for an infinite strip oriented along an arbitrary direction (represented by the vector $\vec{\theta}=\hat{e}_x \cos \theta + \hat{e}_y \sin \theta$, $\theta \in [0,\pi]$) within the $x$-$y$ plane of a $4$-dimensional homogeneous spacetime (see Fig. \ref{fig:plotshowrotate}). 
The generic anisotropic spacetime is,
\begin{equation}\label{hom}
    d s^{2}=g_{t t} d t^{2}+g_{z z} d z^{2}+g_{x x} d x^{2}+g_{y y} d y^{2}.
\end{equation}

To streamline our analysis, we introduce a new coordinate system,
\begin{equation}\label{coor}
    \tilde{t}=t, \quad \tilde{z}=z, \quad \tilde{x}=x \cos (\theta)+y \sin (\theta), \quad \tilde{y}=y \cos (\theta)-x \sin (\theta),
\end{equation}
where here $\tilde{x}$ points to $\vec \theta$ since $\vec{\theta}=\hat{e}_{\tilde{x}}$, and now the minimum surface can be described by $\tilde{z}(\tilde{y})$ as it is invariant along $\tilde{x}$. In the coordinate system \eqref{coor}, the induced metric on the hypersurface $\tilde{z}(\tilde{y})$ at $\tilde{t}$= const can be expressed as
\begin{equation}\label{indu}
    d \hat{s}^{2} =g_{\tilde{x} \tilde{x}} d \tilde{x}^{2}+\left(g_{z z} z^{\prime}(\tilde{y})^{2}+g_{\tilde{y} \tilde{y}}\right) d \tilde{y}^{2}+g_{\tilde{x} \tilde{y}} d \tilde{x} d \tilde{y}.
\end{equation}
For simplicity, we omit the common factors of the area of the minimum surface, so the expression for HEE can be summarized as:
\begin{equation}\label{hee1}
    S=\int_{\Sigma} \mathcal{L} d \tilde{y},
\end{equation}
where $\mathcal{L}=\sqrt{\hat{g}}=\sqrt{g_{x x} g_{y y}+g_{z z} z^{\prime}(\tilde{y})^{2}\left(g_{x x} \cos ^{2}(\theta)+g_{y y} \sin ^{2}(\theta)\right)}$. This expression serves as the Lagrangian density for the minimum surface with angular dependence, and thus the anisotropic HEE can be obtained by solving the EoM derived from this Lagrangian density.

\begin{figure}[htbp]
    \includegraphics[height=0.2\textwidth]{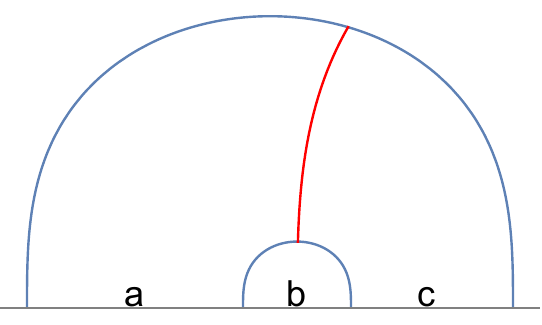}
    \caption{Visualization of the minimum cross-section (red curve) that cuts through the entanglement wedge.}
    \label{fig:stripe3}
\end{figure}

Entanglement entropy is a suitable measure of entanglement for bipartite pure quantum states, quantifying the degree of entanglement by the Von Neumann entropy of one of the subsystems. However, it is unreliable for mixed quantum states, which can contain a combination of quantum and classical correlations. To assess entanglement in mixed states, other measures are necessary, such as the Entanglement of Purification (EoP), proposed as a novel measure for mixed-state entanglement. It involves a double minimization procedure to purify the additional degrees of freedom associated with mixed states \cite{Terhal:2002zz}. The holographic EoP is proposed to be proportional to the area of the minimum cross-section of the entanglement wedge (EWCS) \cite{Takayanagi:2017knl}. When the widths of the two subregions and their separation are $a$, $c$ and $b$ respectively, we label this configuration with $(a, b, c)$  (as illustrated in FIG. \ref{fig:stripe3}). 

We adopt computational techniques outlined in \cite{Liu:2020blk} to compute EWCS, a methodology applied across various models in our studies. The minimum cross-section is parametrized by $z$ for convenience, and the Lagrangian density is thus expressed as
\begin{equation}\label{eq,ll2}
    \mathcal{L}_{\text{EWCS}}=\sqrt{g_{x x}(z) g_{y y}(z) y^{\prime}(z)^{2}+g_{z z}(z) \left(g_{x x}(z) \cos ^{2}(\theta)+g_{y y}(z) \sin ^{2}(\theta)\right)}.
\end{equation}
Solving the EoM derived from this Lagrangian density for the EWCS with angular dependence, we can capture the anisotropy in mixed-state entanglement within the holographic model.

\subsection{The Butterfly Velocity}\label{butterfly}

The butterfly effect, also known as quantum chaos, refers to the exponential growth of the Hermitian operator commutator triggered by a small initial perturbation \cite{Blake:2016wvh,Ling:2017jik,Baggioli:2018afg,Roberts:2014isa,Maldacena:2015waa,Polchinski:2015cea}. This intriguing phenomenon has been the subject of extensive research within the domain of holographic theories, particularly for its significant application in diagnosing the superconductivity phase transitions and quantum phase transitions \cite{Blake:2016wvh,Ling:2017jik,Baggioli:2018afg}. The ubiquity of the butterfly effect in holographic theories stems from its exclusive dependence on the near-horizon data of the gravitational bulk theory. The butterfly velocity $v_B$, quantifying the butterfly effect, can be extracted from the shockwave solution near the horizon and characterizes the speed of information propagation in a quantum system \cite{Roberts:2016wdl}.

For a generic anisotropic black brane \eqref{metric}, the butterfly velocity also exhibits anisotropic characteristics. Denoted as $\bar{v}_{B}(\theta)$, it is expressed as \cite{Ling:2017jik}:
\begin{equation}\label{vb0}
    \bar{v}_{B}(\theta)=\left.v_{B} \sqrt{\cos ^{2}(\theta)+\frac{V_{1}(z) \sin ^{2}(\theta)}{V_{2}(z)}}\right|_{z=1},
\end{equation}
where $v_{B}=\bar{v}_{B}(0)$ represents the butterfly velocity along the $x$-direction. It can be explicitly computed as
\begin{equation}\label{vb1}
    v_{B}=\left.\sqrt{\frac{-2 \pi \hat{T} V_{2}(z)}{V_{2}(z)\left[V_{1}^{\prime}(z)-2 V_{1}(z)\right]+V_{1}(z)\left[V_{2}^{\prime}(z)-2 V_{2}(z)\right.}}\right|_{z=1},
\end{equation}
where the prime denotes the derivative with respect to the coordinate $z$. The $ v_B(\theta) $ exhibits rotational symmetry with a periodicity of $ \pi $ and mirror symmetry about $ \theta = \frac{\pi}{2} $. Therefore, we study $v_B$ as a function of temperature and $\theta\in [0,\pi/2]$.
Examining formula \eqref{vb0}, the butterfly speed in this model exhibits a direct relationship with the angle $\theta$. Furthermore, the butterfly velocity is solely influenced by $V_{1}(z)$ and $V_{2}(z)$ at IR, so the butterfly velocity is only related to the geometry at the horizon of this system. The anisotropy of butterfly velocity is determined only by $V_1(1)$ and $V_2(1)$, hence the butterfly velocity is isotropic at $T_{II}$.

In the subsequent section, we will provide detailed explorations of these holographic information-related quantities to uncover how dynamical and static quantities get impacted by anisotropy.

\section{The Study of Anisotropy in Quantum Information}\label{sec:computation}
Due to the inherent anisotropy in this model, physical quantities can exhibit angular dependence in the $x$-$y$ plane. To discuss anisotropy explicitly and conveniently, we employ a method of variance to quantify the degree of anisotropy in values of physical quantities. The anisotropy $\mathcal{V}(Q)$ of a physical quantity $Q$ in this model is computed as 
\begin{equation}\label{variance}
    \mathcal{V}(Q) =\frac{1}{\pi /2} \int_{0}^{\pi /2}\left[Q(\theta)-\bar{Q}\right]^{2} d \theta.
\end{equation}
A larger variance indicates more significant anisotropy.

\subsection{The Anisotropy in Holographic Entanglement Entropy}\label{sec:HEE}

To investigate the anisotropy in HEE, we need to locate the factors that might influence it. Firstly, the width of the subregion under consideration plays a crucial role in determining how far the minimum surface can extend. By varying the width, we can effectively probe the anisotropy at different energy scales, allowing us to explore its behavior across different regimes. Secondly, the temperature of the background spacetime is another important factor that can affect the anisotropy of HEE. As the temperature drops below $T_{II}$, an isotropic point emerges in the background spacetime, which in turn impacts the anisotropy pattern of HEE. Therefore, to gain a comprehensive understanding of the anisotropy of HEE, we systematically explore the influence of two key parameters: the width of the subregion (denoted as $w$) and the temperature $T$ of the system.

\begin{figure}
    \centering
    \includegraphics[height=0.3\textwidth]{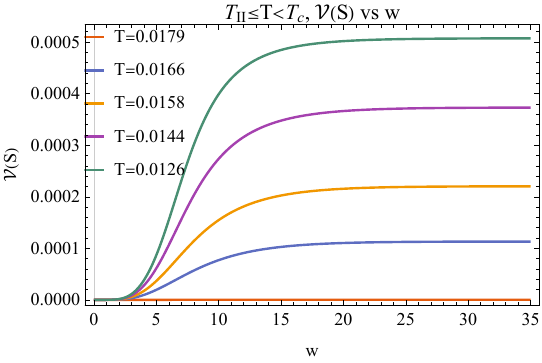}
    \includegraphics[height=0.3\textwidth]{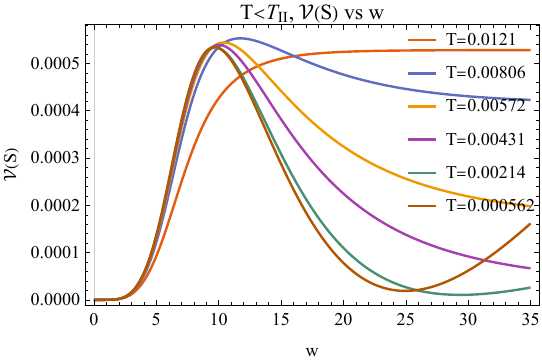}
    \caption{The anisotropy of HEE changes with width at varying temperatures.}
    \label{fig:hee_anisocase1}
\end{figure}

We show HEE behaviors with width in Fig. \ref{fig:hee_anisocase1}. 
For subregions with a small width (see the small region in the lower left corner of both plots in Fig. \ref{fig:hee_anisocase1}), HEE is almost isotropic at any temperature. This is because the small minimum surface is always located near the boundary, where $V_{1}$ and $V_{2}$ are nearly equal.

Next, we study the situations where the subregion's width is relatively large. In this case, the anisotropy of HEE will be more evident because the minimum surface stretches into a deeper region where anisotropy is more pronounced. Since the anisotropy is affected by temperature, leading to transitions to isotropy at certain scales, the HEE will act differently at temperatures above and below $ T_{II} $. Thus, we will concentrate on two temperature ranges: the first is $T_{II} < T < T_c$ (the left plot in Fig. \ref{fig:hee_anisocase1}) and the second is $ 0 < T < T_{II} $ (the right plot in Fig. \ref{fig:hee_anisocase1}). 

\begin{figure}[htbp]
    \includegraphics[height=0.22\textwidth]{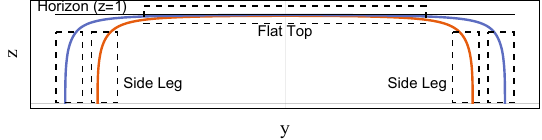}
    \caption{HEE for large subregions. The black brane horizon is located at $z=1$. The two primary contributions to HEE, namely Flat Top and Side Leg, are highlighted in the figure.}
    \label{fig:largehee}
\end{figure}

\begin{figure}
    \centering
    \includegraphics[height=0.2\textwidth]{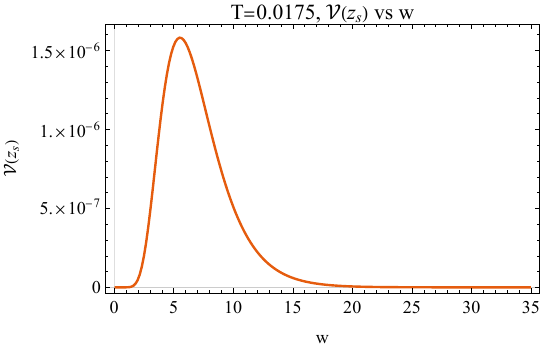}
    \includegraphics[height=0.2\textwidth]{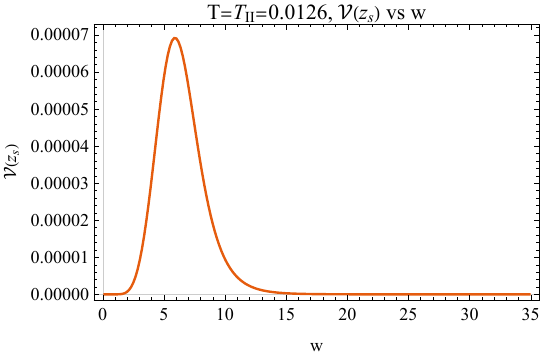}
    \includegraphics[height=0.2\textwidth]{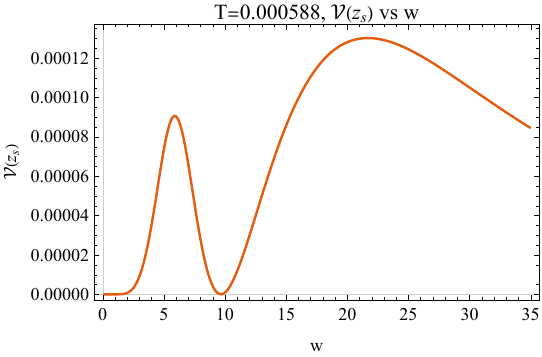}
    \caption{From left to right are the anisotropy of $z_s$ changes with width at $T>T_{II}$, $T_{II}$ and $T<T_{II}$ respectively.}
    \label{fig:Vzsvsw1}
\end{figure}

In the temperature range $T_{II} < T < T_c$, as the width increases, the anisotropy of HEE initially rises from zero and eventually stabilizes. The initial gradual accumulation of anisotropy can be explained by the rapid extension of the minimum surface into the anisotropic region as the width increases. As the minimum surface becomes significantly large and approaches the horizon (see Fig. \ref{fig:largehee}), its contributions to HEE can be primarily decomposed into two segments,
\begin{enumerate}
    \item Flat Top: The flat top region, close to the horizon, analogous to the flat surface of an overturned cup. We mark the radial position of this flat region as the radial coordinate of the turning point or top of the minimum surface $z_s$, which is close to $1$ as the minimum surface approaches the horizon. The area of this segment is proportional to the thermodynamic entropy $S_{Th}$. Notably, since $S_{Th}=s \cdot w$, where $s=\frac{\sqrt{V_1 \left( 1 \right) V_2 \left( 1 \right)}}{\mu^{2}}$ denotes the dimensionless entropy density, the contribution of the Flat Top is isotropic.
    \item Side Leg: The side structures connecting this flat area to the boundary, analogous to the upright legs of a cup. The area of this segment is anisotropic.
\end{enumerate}
When the width is large, HEE usually behaves like thermodynamic entropy. The Flat Top part is the main factor in HEE, but the anisotropy of HEE mostly comes from the Side Leg part. This happens because, for large widths, the Flat Top stays almost level and at the same position on $z$-axis for different angles when the temperature is between $T_{II}$ and $T_c$, as shown in the left and the middle plot in FIG. \ref{fig:Vzsvsw1}.
As a result, when the width increases further, the anisotropy in HEE becomes mainly determined by the side leg parts, leading to a stabilizing pattern of anisotropy.

When the temperature decreases below $T_{II}$, the anisotropy of HEE initially increases, followed by a subsequent decrease with width, and it will reaccumulate as the temperature is sufficiently low (see the right plot in Fig. \ref{fig:hee_anisocase1}). To further investigate this subsequent decreasing behavior in low-temperature systems, we study the variance of the anisotropy in $z_s$ (the radial coordinate of the top or turning point of the minimum surfaces) at different temperatures (see Fig. \ref{fig:Vzsvsw1}). Comparing the three plots in Fig. \ref{fig:Vzsvsw1}, we observe that $z_s$ of large surface exhibits much more significant variations at low temperatures than that of the case at temperatures above $T_{II}$. This behavior can be attributed to the emergence of the isotropic point at $T_{II}$, which represents a critical shift in the system's anisotropic behavior. The presence of the isotropic point at $T_{II}$ signifies the recovery of isotropy at this IR scale. As the temperature decreases further, this isotropic scale shifts into the bulk and significantly impacts the background geometry. This shift in isotropy scale has a direct effect on the position of the Flat Top segment $z_s$. Therefore, this can again impact the anisotropy of HEE when $T<T_{II}$, counteracting the anisotropy generated by the Side Leg segment. In summary, it is the isotropic point that drives the decreasing behaviors of HEE at low temperatures.

\begin{figure}
    \centering
    \includegraphics[height=0.3\textwidth]{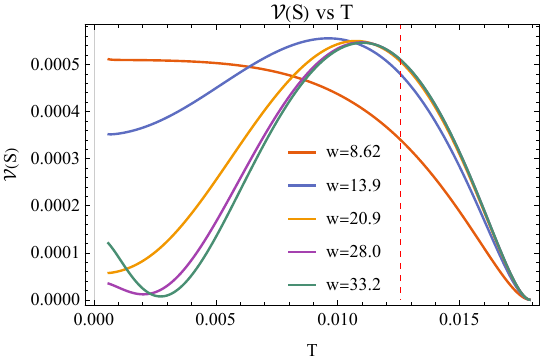}
    \caption{The anisotropy of HEE changes with temperatures. The red dashed line represents the isotropic temperature of IR $T_{II}$.}
    \label{fig:heevsT}
\end{figure}
\begin{figure}
    \centering
    \includegraphics[height=0.3\textwidth]{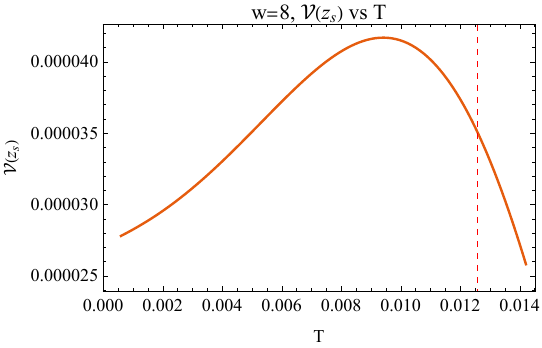}
    \includegraphics[height=0.3\textwidth]{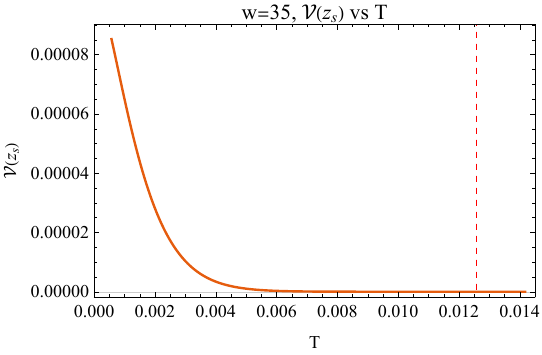}
    \caption{The anisotropy of $z_s$ changes with temperature when the subregion's width is small and large. The red dashed line denotes the isotropic temperature of IR $T_{II}$.}
    \label{fig:VzsvsT}
\end{figure}

Next, we delve into how the anisotropy of HEE changes with temperature (see Fig. \ref{fig:heevsT}). When $T_{II}< T < T_{c}$, the anisotropy of HEE increases with decreasing temperature. This behavior can be attributed to the rapid growth of spacetime anisotropy with decreasing temperature when $T \lesssim T_{c}$. However, when $T < T_{II}$, HEE can detect the emergence of isotropic points in the energy scale, leading to a decrease in the anisotropy of HEE. The HEE of larger subregions experiences a steeper decrease compared to smaller ones, indicating that the larger minimum surfaces can probe more anisotropy near the horizon. From a physical perspective, this observation suggests that the HEE of a larger subregion involves more information about the anisotropy of IR. Therefore, the HEE of larger subregions could be a better method to characterize the transition of anisotropy in this system. We have discovered that the HEE for large subregions can identify the temperature at which the isotropic point emerges at the IR.
\footnote{As a similar phenomenon, the maximum anisotropy of the $c$-function of the large subregion can also identify the quantum critical point, as shown in FIG. 3 of reference \cite{Baggioli:2020cld}. }
This is surprising because, for wide subregions, the HEE is primarily influenced by the thermal entropy, which is isotropic. As discussed earlier, the local maximum of $\mathcal V(S)$ at the transition temperature $T_{II}$ is due to the maximum anisotropy contributed by the Side Leg. Therefore, we conclude that the HEE can determine the temperature at which the isotropic point emerges at the IR, but it is not solely the property of the IR itself that governs this phenomenon (in previous studies, properties of HEE are determined by the thermal entropy); rather, it is the Side Legs that depend on the entire content of the bulk geometry. Moreover, when the subregion is large enough, the anisotropy of HEE exhibits reaccumulation as the temperature further decreases. This behavior exhibits a positive relationship with the increasing of $ \mathcal{V}(z_s)$ as the temperature decreases, as illustrated in the right plot of Fig. \ref{fig:VzsvsT}. 

We have conducted a comprehensive analysis of the anisotropy in HEE concerning the subregion's width and temperature. Our findings reveal that the anisotropy of HEE stabilizes with the increase of width in the absence of an isotropic point in spacetime. However, as the temperature decreases, the emergence of the isotropic point significantly impacts the anisotropic behavior of HEE, especially for large subregions. To further understand how isotropy influences mixed-state entanglement, we will next examine the anisotropy in EWCS.

\subsection{The Anisotropy in Entanglement Wedge Cross-section}\label{sec:eop_phenomena}

As explained in Section \ref{sec:HEE}, the subregion configuration $(a,b,c)$ and temperature can influence the anisotropy in EWCS. We will elucidate their effects on the anisotropy in EWCS within this model.

\begin{figure}
    \centering
    \includegraphics[height=0.3\textwidth]{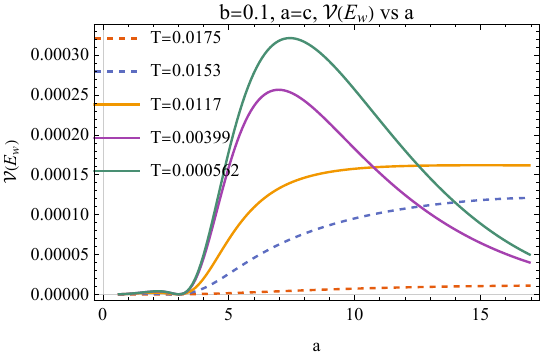}
    \caption{The anisotropy of EWCS of symmetric configuration changes with $a$ and $c$. The dashed lines represent the temperature range $T_{II} < T < T_c$, and the solid lines represent temperatures below $T_{II}$.}
    \label{fig:Veopvsa}
\end{figure}

We first investigate the relationship between $\mathcal{V}(E_W)$ and the configuration $(a, b, c)$. Specifically, we focus on a symmetric configuration ($a=c$) and examine how the anisotropy of EWCS changes when we keep $b$ fixed and increase the sizes of subregions $a$ and $c$ simultaneously. The results are presented in Fig. \ref{fig:Veopvsa}. When $T_{II} < T < T_c$, the anisotropy of EWCS shows a small peak followed by a continuous increase as $a$ and $c$ increase (see the dashed lines in Fig. \ref{fig:Veopvsa}). However, when $T<T_{II}$, a different pattern emerges (see the solid lines in Fig. \ref{fig:Veopvsa}). Initially, the minor peak in the anisotropy of EWCS persists as $a$ and $c$ increase, but further increases in $a$ and $c$ lead to a second peak in the EWCS before it starts to decrease again.

\begin{figure}
    \centering
    \includegraphics[height=0.35\textwidth]{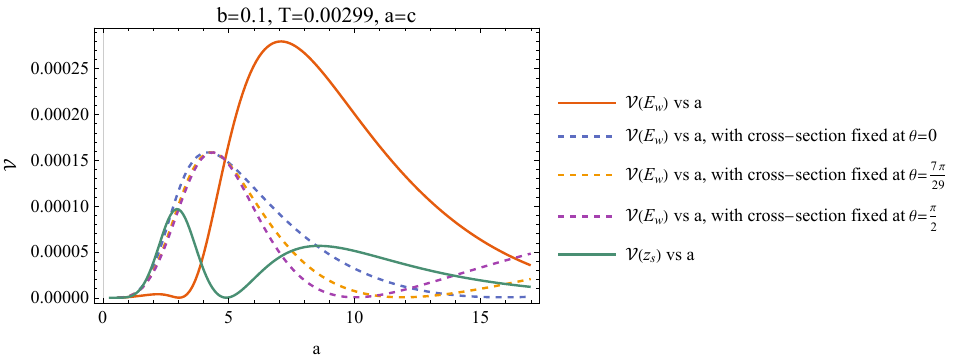}
    \caption{The figure presents the anisotropy of EWCS, the anisotropy of EWCS of cross-section fixed at three specific angles, and the anisotropy of $z_s$ with changing $a$ and $c$ at a low temperature.}
    \label{fig:Veopvsath}
\end{figure}

To gain deeper insights into the anisotropy of EWCS with respect to increasing subregions' sizes $a$ and $c$, particularly focusing on the emergence of two peaks, we adopt a stratified approach. Specifically, we decouple the influences in anisotropy and divide the analysis into two parts. First, We solely examine the variation in the integration by keeping the minimum surfaces and the shape of the corresponding cross-section constant at a specific value of $\theta$. Next, we delve into the influence of the cross-section's shape by examining the variance in $z_s$. By differentiating between these two scenarios, we can systematically assess the impact of each: the former representing the direct influence of integration under the anisotropic spacetime background, and the latter capturing the higher-order effects that stem from the geometric variation of cross-sections. We examine the anisotropy of EWCS when the cross-section is fixed at three representative angles (as indicated by the dashed lines in Fig. \ref{fig:Veopvsath}).
In this variance of the integration, the minor peaks are consistently absent, which is qualitatively similar to anisotropy in HEE with respect to increasing width (Fig. \ref{fig:hee_anisocase1}). This alignment can be explained based on the earlier discussion in Section \ref{sec:HEE}. When the subregion's width is large, the Side Leg segment of a large minimum surface is vertical, and similarly, the cross-section of entanglement wedge in the symmetric configuration is also vertical due to this symmetry. Therefore, the anisotropy of the latter also resembles the former. On the other hand, the variation of the shape of this cross-section can be reflected by the $z_s$ of the larger minimum surface with width $w=a+b+c$. The peaks of the anisotropy in EWCS match well with the peaks in the anisotropy of $z_s$ (as shown by the orange line and green line in FIG. \ref{fig:Veopvsath}). This suggests that, instead of being influenced by the integration of cross-sections, the peaks in the anisotropy of EWCS are caused by the strong anisotropy in the $z_s$ of the larger minimum surface in the entanglement wedge, i.e. the anisotropy in the shape of cross-sections.

However, the question remains: why does not HEE exhibit the same minor peak as EWCS? From the perspective of dual geometry, the anisotropy in EWCS is mainly influenced by the anisotropy near the energy scale $z_s$ (i.e., the position of the top of the cross-section), since the cross-section of symmetric configurations is a vertical plane connecting the top of two surfaces. Therefore, significant anisotropy in $z_s$ leads to notable anisotropy in EWCS. This significant impact of $z_s$ on EWCS differs from its impact on HEE. This is because HEE is influenced by the entire minimum surface and encompasses all geometric information from its top to the AdS boundary, making it less sensitive to the energy scale $z_s$ than EWCS. Consequently, the anisotropy of HEE does not match well with the peaks of anisotropy in $z_s$ like EWCS. On the other hand, from the perspective of the boundary dual quantum field theory, entanglement entropy has two contributions: the leading isotropic thermal contribution and the subleading quantum effects \cite{Fischler:2012uv}. The quantum contribution is subordinated by the isotropic thermal contribution. The mixed-state entanglement measured by EWCS of two subregions can cancel out the thermal contribution \cite{Huang:2019zph}, thereby better capturing the properties of anisotropy. 

\begin{figure}
    \centering
    \includegraphics[height=0.3\textwidth]{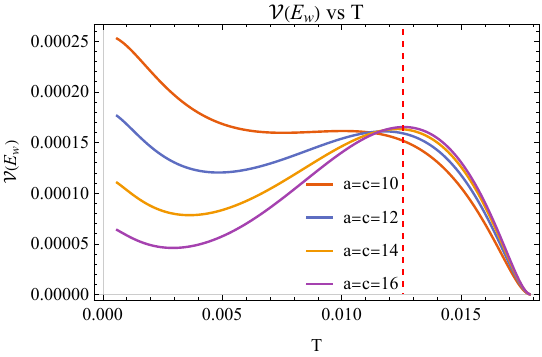}
    \caption{The anisotropy of EWCS of symmetric configuration changes with temperature. The red dashed line denotes the isotropic temperature of IR $T_{II}$.}
    \label{fig:VeopvsT}
\end{figure}

Finally, we examine how the anisotropy of EWCS changes with temperature (see FIG. \ref{fig:VeopvsT}). As the temperature decreases near $T_c$, the spacetime anisotropy increases quickly, causing a sharp rise in $\mathcal{V} (E_W)$. However, once the temperature falls below $T_{II}$, the anisotropy in EWCS of the larger configuration decreases, while that of the smaller configuration continues to increase. Furthermore, as the temperature decreases even further, the anisotropy of any configuration increases consistently. Notably, as the configuration gets larger, the extreme point of $\mathcal{V} (E_W)$ approaches closer to the isotropic temperature of IR $T_{II}$, similar to the pattern of HEE in FIG. \ref{fig:heevsT}. From the perspective of dual geometry, the emergence of the isotropic point represents the transition of anisotropy near the horizon. As mentioned in the previous paragraph, the anisotropy of EWCS is mainly determined by the anisotropy near $z_s$. As the sizes of the subregions increase, the cross-section can extend further into the IR region, thereby enhancing the sensitivity to probe the anisotropy in that area. As a result, the extreme point in EWCS matches the $T_{II}$ more precisely (as shown in FIG. \ref{fig:VeopvsT}). Therefore, EWCS serves as a highly sensitive probe for detecting the isotropic point and the transition of anisotropy.

In summary, this extensive study of the anisotropy of EWCS for symmetric configuration shows similar behavior to HEE. However, EWCS, with its capability of measuring mixed-state entanglement, is more sensitive to the isotropic point than HEE. Having discussed two static information-related quantities thus far, we will now shift our focus to a dynamic information-related quantity—the butterfly velocity.

\subsection{The Anisotropy in Butterfly Velocity}

The butterfly velocity $\bar v_B$ measures how quickly a disturbance or a perturbation spreads its influence across a quantum system, with the relationship informed by the principles of chaos and correlations. The butterfly velocity is determined by the near-horizon geometry, in contrast to HEE and EWCS, which can be significantly influenced by the anisotropy of the bulk. Now, we study how the butterfly velocity changes with the angle $\theta$ and temperature $T$ in this model.

The monotonicity of the butterfly velocity $\bar v_B(\theta)$ with  $\theta$ can be directly determined from its explicit expression \eqref{vb0}. Specifically, for $T_{II} < T < T_c$, which implies $V_1(1) < V_2(1)$, $\bar v_B$ increases monotonically with $\theta$. Conversely, for $T< T_{II}$, which implies $V_1(1) > V_2(1) $, $\bar v_B$ decreases monotonically with $\theta$. Unlike this transition from monotonically increasing to monotonically decreasing observed in $\bar v_B$, other quantities like the HEE or EWCS do not show this monotonic change behavior because they involve the whole bulk geometry, rather than the near horizon region. 
At the specific temperature $T_{II}$, the direction of the maximum $\bar v_B$, which represents the fastest direction of quantum information transmission, undergoes a $90$\textdegree{ }twist. Previous studies have established a connection between the quantum entanglement of a system and its transport properties\cite{Burkard:2003lb,Verstraete:2004ev}. This suggests that the anisotropy of quantum information in a system could be related to the anisotropy of transport. Thus, this $90$\textdegree{ }twist of the fastest $\bar v_B$ is reminiscent of a similar phenomenon observed in Weyl semimetals, where the maximum direction of transportation of charged degrees of freedom, also rotates by $90$\textdegree{ } between the top and bottom of the materials \cite{Quirk:2023}. 

\begin{figure}
    \centering
    \includegraphics[height=0.32\textwidth]{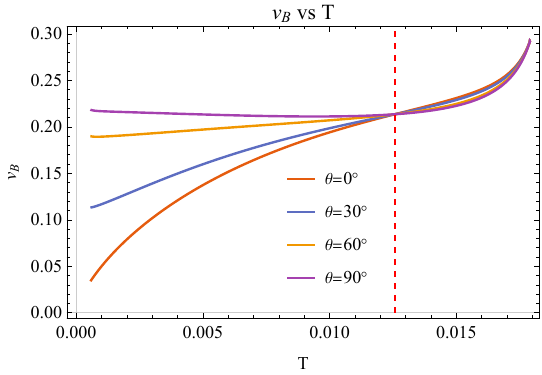}
    \caption{Variations in butterfly velocity with temperature at different angles. The red dashed line denotes the isotropic temperature of IR $T_{II}$.}
    \label{fig:vbvsT}
\end{figure}

Now, we examine how butterfly velocity changes with temperature and present the results in FIG. \ref{fig:vbvsT}. A notable observation is that an intersection point emerges at $T = T_{II}$, which represents the aforementioned transition of the monotonicity of $\bar v_B$ with $\theta$. From a dual geometry perspective, butterfly velocity is determined by the geometry at the horizon. Therefore, the transition of monotonicity of $\bar v_B (\theta)$ directly reflects the transition of anisotropy of spacetime at the IR. From the left segment of FIG. \ref{fig:vbvsT}, it is evident that $\bar v_B$ exhibits increasing variation across different angulars as the temperature decreases, so $\mathcal{V}(\bar v_B)$ increases with decreasing temperature. The increase in anisotropy of $\bar v_B$ implies the increase in horizon anisotropy below the transition point $T_{II}$. This is supported by the significant reaccumulation of the anisotropy in EWCS of large configurations at low temperatures (as shown in FIG. \ref{fig:VeopvsT}). From a perspective of dual quantum field theory, $\bar v_B$ captures the large-scale degrees of freedom. Therefore, $\bar v_B$ can also serve as a tool to characterize the transition of anisotropy at the IR in this system.

In summary, as an information-related quantity defined by the geometry at the horizon, butterfly velocity exhibits a transition of monotonicity crossing the certain point $T_{II}$. It serves as a good tool to probe the anisotropy of quantum information at the IR energy scale.

\section{Discussion}\label{sec:discuss}

In this work, we have conducted a comprehensive investigation into the anisotropic behavior of the holographic p-wave superconductor system and various information-related quantities, including HEE, EWCS, and butterfly velocity, in this system. Our key findings and main conclusions are as follows:

\begin{itemize}

    \item The holographic p-wave superconductor model exhibits anisotropy due to the presence of the vector field, which explicitly breaks isotropy in the $x$-$y$ plane. At temperature $T_{II}$, an isotropic point emerges in the bulk geometry where $V_1=V_2$ at the horizon. The existence of isotropic point is semi-analytically proved through studying the anisotropy of UV and IR regions near critical point $T \lesssim T_c$ and the zero-temperature limit $T \to 0$.
    \item Static information-related quantities such as HEE and EWCS are sensitive to the anisotropy in the bulk geometry. When the sizes of the subregions are large, they behave similarly but show notable differences as the subregion size changes due to their distinct definitions within the dual gravity system. In the symmetric configuration, the cross-section resides entirely inside the bulk, connecting the two minimum surfaces. Meanwhile, the minimum surfaces of the HEE connect the boundary and stretch into the horizon. Therefore, EWCS is more sensitive to the anisotropy near the IR region than HEE. The butterfly velocity, as a dynamical information-related quantity, is determined by the horizon geometry and can serve as a precise probe to the anisotropy at the IR region.
    \item All these information-related quantities can capture the emergence of isotropic points at $T_{II}$, but they differ in how they diagnose it. Specifically, the butterfly velocity shows a change in monotonicity at the temperature $T_{II}$, which directly reflects the change in isotropy at the IR. Both EWCS and HEE of large configurations approach their extreme points near $T_{II}$ due to the sensitivity of the energy scale near the IR region. The maximum for EWCS is very close to $T_{II}$, while the maximum for HEE is less so compared to it. Consequently, EWCS and butterfly velocity are better choices for characterizing the emergence of isotropic points.
          
\end{itemize}
In summary, the distinct anisotropic behavior informs us of the sensitivity of different probes to various bulk regions.

The holographic p-wave superconductor model has a unique feature in its transition of anisotropy compared to other models like the Q-lattice dilaton model and the EMDA Q-lattice model \cite{Ling:2016wyr, Fu:2022qtz,Donos:2013eha,Donos:2014uba}. This uniqueness comes from the intrinsic symmetry-breaking mechanisms in this model. Instead of using a spatially dependent field to achieve anisotropy like the Q-lattice model, the p-wave superconductor is achieved by introducing a vector field with a non-vanishing $ x $ component to explicitly break isotropic symmetry, which is key to introducing strong enough deformations so as to recover isotropy at a certain scale. This causes the emergence of isotropic points, resulting in a $90$\textdegree{ } rotation in the fastest direction of quantum information transmission. Additionally, from the perspective of renormalization group flow, there exists a significant transition between anisotropy to isotropy at a certain scale $ z $, which can translate to a certain size or scale of a system. Interestingly, similar findings related to the relationship of anisotropy and size have been observed in condensed matter systems like turbulent flows, magnetic nanoparticles and two-dimensional hexagonal boron nitride, some of which also employed the renormalization group method \cite{Biferale2005ai,Carati1989rg,morup2007experimental,pacakova2016understanding,thomas2016directional}.

At the end, we point out several worthwhile research directions that remain open for future exploration. First, extending our computations to other holographic anisotropic models, such as the d-wave superconductor model and other alternate p-wave superconductor models can further examine the generality of the emergence of the isotropy in anisotropy systems. This will provide valuable evidence for the conclusions presented in this work. Moreover, inhomogeneous lattice models can also exhibit anisotropy, which we believe can display more strongly deviation from the homogeneous lattices model \cite{Ling:2015ghh,Blake:2013owa,Ling:2013nxa,Ling:2013aya,Rao:2020hall}. In addition, it is desirable to explore an analytical understanding on the explicit condition for $V_1 = V_2$ at the horizon at a specific temperature in this model. This could offer a more complete analytical understanding of our results. Also, probing the anisotropy by using other dynamical quantities, particularly transport coefficients like thermal and electrical conductivities, could unveil additional facets of the anisotropy structures in strongly coupled anisotropic systems. 

\section*{Acknowledgments}

We thank Run-qiu Yang for helpful discussions. Mu-jing Li expresses gratitude for the unwavering support and encouraging words provided by her parents throughout this endeavor. Peng Liu would like to thank Yun-Ha Zha and Yi-Er Liu for kind encouragement during this work. This work is supported by the Natural Science Foundation of China under Grant No. 11905083, as well as the Science and Technology Planning Project of Guangzhou (202201010655).

\appendix

\section{Numerical Methods and Technical Considerations}
\label{app:technical}

The equations of motion (EoMs) for the p-wave model yield a system of highly non-linear, second-order differential equations. To solve these equations numerically, we employed a combination of spectral methods and iterative techniques \cite{boyd:2001cfs}. Specifically:

\subsection{Discretization and Iteration}

\begin{enumerate}
    \item We discretized the radial coordinate $z$ using Chebyshev collocation points.
    \item The resulting system was solved using the Newton-Raphson method.
\end{enumerate}

For the solutions presented in FIG. \ref{fig:V1V2q15} and \ref{fig:TcTII}, we used $30$ collocation points and iterated until the residual error fell below $10^{-6}$. The error is computed as the L2 norm of the residuals of the EoMs evaluated at the $30$ collocation points. To ensure the reliability of our numerical results, we conducted convergence tests by increasing both the number of collocation points and the precision of numerics. These tests confirmed that our results are indeed convergent. Through this process, we determined that $30$ collocation points provide an optimal balance between accuracy and computational efficiency for the problems at hand.

\subsection{Low-Temperature Considerations}

As the temperature approaches zero, the solutions exhibit near-singular behavior in the vicinity of the horizon. To mitigate numerical instabilities, we implemented a coordinate stretching technique:
\begin{equation}
    z \rightarrow \tanh^{-1}(\tanh(2)z)
\end{equation}
This transformation enhances the resolution near the horizon, improving the stability of the iteration process.

\begin{figure}[ht]
    \centering
    \includegraphics[width=0.8\textwidth]{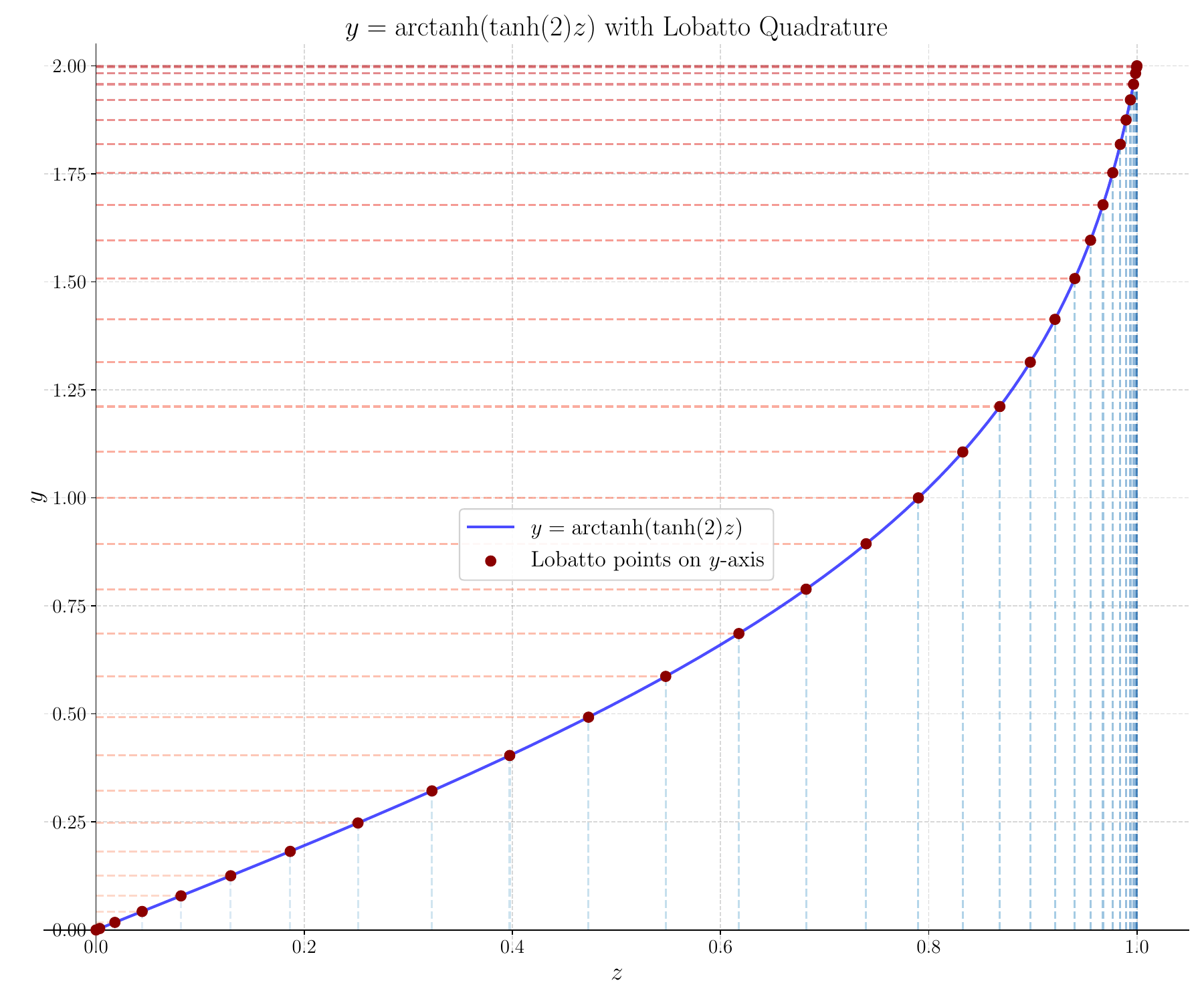}
    \caption{Configuration of the collocation points and the coordinate stretching transformation.}
    \label{fig:lobatto_points}
\end{figure}

Fig. \ref{fig:lobatto_points} demonstrates the configuration of the collocation points and the effect of the transformation $z \rightarrow \tanh^{-1}(\tanh(2)z)$, which enhances the resolution near the horizon, thereby improving the stability of the iteration process.

\subsection{Derived Quantities}

The numerical results for the holographic HEE, EWCS, and butterfly velocity were computed based on these background solutions.

\subsection{Einstein-Maxwell-Dilaton-Axion (EMDA) Model}

For the EMDA model \cite{Donos:2014uba,Fu:2022qtz}, additional numerical challenges arise at low temperatures due to more pronounced near-horizon singularities. To address these issues:

\begin{enumerate}
    \item We increased the number of collocation points to 120.
    \item We employed a more aggressive stretching function:
          \begin{equation}
              z \rightarrow \tanh^{-1}(\tanh(6)z)
          \end{equation}
    \item We enhanced the numerical precision to \texttt{3*MachinePrecision}.
\end{enumerate}

These adjustments were crucial to eliminate numerical artifacts and accurately capture the anisotropic behavior of the system.

All numerical computations were performed using Mathematica, with custom-written modules for spectral differentiation and integration.

\end{document}